\documentclass[a4paper]{article}

\usepackage{INTERSPEECH2019}
\usepackage{xcolor}
\usepackage[utf8]{inputenc}
\usepackage{tikz}
\usepackage{url}

\usepackage{amsmath,graphicx}
\usepackage{lipsum}
\usepackage{url,multirow}
\usepackage{standalone}
\usepackage{booktabs}
\usepackage{pgfplots}
\usetikzlibrary{positioning,chains,shapes.geometric, calc, shadows, shapes.misc}
\ninept
\usepackage[font={small},skip=2pt,margin=10pt]{caption}

\def\thebibliography#1{\begin{center}
{\bf \large References\vspace{-.5em}\vspace{2pt}}
\end{center}\eightpt\list
 {[\arabic{enumi}]}{\settowidth\labelwidth{[#1]}\leftmargin\labelwidth
 \advance\leftmargin\labelsep
 \usecounter{enumi}}
 \def\newblock{\hskip .11em plus .33em minus .07em}
 \sloppy\clubpenalty4000\widowpenalty4000
 \sfcode`\.=1000\relax}

\title{Comparison of Lattice-Free and Lattice-Based \\ Sequence Discriminative Training Criteria for LVCSR}
\name{Wilfried Michel, Ralf Schl\"uter, Hermann Ney}

\address{
  Human Language Technology and Pattern Recognition, Computer Science Department, \\
  RWTH Aachen University, 52056 Aachen, Germany }
 
\email{\{michel,schlueter,ney\}@cs.rwth-aachen.de}

\begin{document}

\maketitle
\vspace{-5mm}
\begin{abstract}
Sequence discriminative training criteria have long been a standard tool in automatic speech recognition for improving the performance of acoustic models over their maximum likelihood / cross entropy trained counterparts. 
While previously a lattice approximation of the search space has been necessary to reduce computational complexity, recently proposed methods use other approximations to dispense of the need for the computationally expensive step of separate lattice creation.

In this work we present a memory efficient implementation of the forward-backward computation that allows us to use uni-gram word-level language models in the denominator calculation while still doing a full summation on GPU. 
This allows for a direct comparison of lattice-based and lattice-free sequence discriminative training criteria such as MMI and sMBR, both using the same language model during training.

We compared performance, speed of convergence, and stability on large vocabulary continuous speech recognition tasks like Switchboard and Quaero.
We found that silence modeling seriously impacts the performance in the lattice-free case and needs special treatment.
In our experiments lattice-free MMI comes on par with its lattice-based counterpart. 
Lattice-based sMBR still outperforms all lattice-free training criteria.

\end{abstract}
\noindent\textbf{Index Terms}: speech recognition, acoustic modeling, sequence discriminative training, lattice-free MMI

\section{Introduction}
\label{sec:intro}
Hybrid hidden Markov models (HMM) \cite{Bourland94} have been used for decades in state-of-the-art large vocabulary continuous speech recognition (LVCSR) systems.
Gaussian hybrid HMMs have since been replaced with feed forward deep neural networks (DNN) \cite{Seide2011} and recurrent long short-term memory neural networks (LSTM)\cite{Hochreiter:1997:LSTM}.
All approaches have a common discrepancy that while they are modeling speech which is sequential data, the training criterion maximum likelihood or cross-entropy is defined on the frame level and minimizes the frame error rate (FER).
However, the recognition is done on sequence-level and also incorporates additional knowledge sources like the language model.
Furthermore, the evaluation criterion is the word error rate (WER).
This mismatch between training and recognition suggests that using training criteria which are based on whole sequences and incorporate additional knowledge sources can be helpful.

Sequence discriminative training of neural networks for Automatic Speech Recognition (ASR) has been shown to provide a significant reduction in word error rates, in contrast to cross-entropy training for each prevalent modeling technique.
Discriminative Gaussian HMMs (GHMMs) achieved best performance when trained using sequence-discriminative criteria like maximum mutual information (MMI)~\cite{mmi}, minimum phone error (MPE)~\cite{povey_phd} or minimum Bayes risk (MBR) related criteria ~\cite{mbr}.
Significant improvements using sequence-discriminative criteria for feed forward networks have been reported in literature~\cite{kingsbury2012, Vesel2013SeqTr}.
These improvements hold true for deep bidirectional LSTM-RNNs (BLSTM), although BLSTMs inherently make use of the whole sequence and are able to incorporate more contextual information~\cite{voigtlaender2015}.

One shortcoming of these criteria is, that for the calculation of the gradients for model updates forward backward computations over the exponentially growing set of all possible HMM state sequences are necessary.
To make these computations feasible, some approximations are necessary.
The traditional way to approximate the search space is to only consider all likely HMM state sequences, which are obtained by decoding the training data with a pretrained model.
The search space is thereby confined to the search lattices which have to be created beforehand and stored to disk. 
Implicit-lattice methods, which do not store lattices, have been explored in the past \cite{Chen06IBM}.

Recently a different, lattice-free approach has been proposed \cite{Povey2016LFMMI}.
Here the search space is not confined to some predetermined state sequences, only the calculation of numerator statistics which encode the supervision information is restricted to loosely follow some fixed, predetermined alignment.
Additionally the language model (LM) has been simplified by limiting the LM context to phone-level four-gram which allows for more frequent recombination of state paths.
This approach has been adopted and adapted to different sequence criteria \cite{Kanda2018LatticefreeSM,Weng2019lfcomparison}, score-fusion and system combination \cite{Kanda2017StudTeach,microsoft-2017-conversational-speech-recognition-system} and settings where we completely dispense with the need of initial GMM models \cite{Povey2018flatstart}.

In this work we aim to get a better understanding of the differences between lattice-based and lattice-free versions of sequence discriminative training criteria. 
We therefore propose two implementations of the forward backward computations which have a lower memory consumption of only $\mathcal{O}(\sqrt{T})$ and $\mathcal{O}(\log(T))$ respectively compared to the $\mathcal{O}(T)$ of a standard implementation \cite{Zweig2000ExactAC,Binder1997fwd-bwd}.
This allows us, at the expense of longer runtime, to dispense of the need of lattices while still avoiding many other approximations used in other lattice-free implementations.
With this we can make use of full length utterances and/or more complex language models.


\section{Lattice-Free vs. Lattice-Based Training}
\label{sec:criteria}
In this work we want to highlight the similarities and the differences between the previously proposed lattice-free, our newly proposed lattice-free, and our lattice-based implementation using the MMI objective function as an example.

On a high level, the MMI criterion is given as the log posterior probability of the reference word sequence $\tilde{w}_1^{\tilde{N}}$ given the audio data $x_1^T$.
A decomposition into similar contributions per training utterance is straight-forward.
\begin{equation}
F_{\mathrm{MMI}} = \log\!\left(\tilde{w}_1^{\tilde{N}} \middle| x_1^T \right)
\end{equation}
Then the usual decomposition into acoustic model and language model is employed and a sequence of time-aligned states is introduced as a hidden variable to arrive at the well-known result
\begin{equation}
\label{eq:mmi}
F_{\mathrm{MMI}} = \frac{\sum\limits_{\{s_1^T \in \tilde{w}_1^{\tilde{N}}\}} p\left(x_1^T,s_1^T\middle|\tilde{w}_1^{\tilde{N}}\right)p\left(\tilde{w}_1^{\tilde{N}}\right)}{\sum\limits_{N,\{ w_1^N \}} \sum\limits_{\{ s_1^T \in w_1^N \}} p\left(x_1^T,s_1^T\middle|w_1^N\right)p\left(w_1^N\right)}
\end{equation}
Where $\{w_1^N\}$ means the set of all possible word sequences of any length and $\{s_1^T \in w_1^N\}$ means the set of all HMM state sequences that are compatible with the given word sequence.
Now different schemes apply different approximations.

\subsection{Lattice-Based Training}
In the classical, lattice-based approach, the sums in the denominator are approximated. 
Instead of summing over all possible word sequences the sum is restricted to a set of likely word sequences which are determined via a decoding run on the training data. 
An utterance specific lattice of hypotheses containing the word identities and word boundary times is written to disk and loaded every time this specific utterance is encountered during the training.

At training time, all possible state sequences for each word arc are constructed and aligned within the given word boundary times.
For each hypothesis word arc the best state path is chosen.
This can be seen as a Viterbi approximation in Equation \ref{eq:mmi}, but formally this is incorrect since a single word sequence can appear with multiple state paths given that they have different word boundary times.

The probabilities can then be modeled by choice, e.g. use an n-gram count-based language model and the usual 3-state HMM topology.

\subsection{Memory-Efficient Lattice-free Training}
In the previously proposed approach for lattice-free MMI no utterance specific restrictions of the denominator sums apply.
Instead, the search space is reduced by constraining the language model context to a phoneme n-gram and simplifying the HMM topology.

In \cite{Povey2016LFMMI}, the sum in the numerator is constrained to loosely follow the reference GMM alignment up to a tolerance of a few frames.
In \cite{povey_2018_slt}, this constrained was relaxed.
Here, the alignment only indirectly influences the state sequences, as the training utterances were cut into chunks with boundaries determined from the initial alignment.

In this work we aim at a direct comparison of lattice-free and lattice-based sequence discriminative training and intend to introduce as few constraints and approximations as possible. 
We therefore keep the original 3-state HMM topology with self loops and skips and also use a word-based language model (unigram). 
There is no restriction on or modification of the length of the training utterances.

This naturally leads to an increased search space, which directly translates to an increased usage of GPU resources.
In a straightforward implementation of the forward backward computations one would do one forward pass while storing the $\alpha$ values to memory and one backwards pass while loading the $\alpha$ values and only keeping the current $\beta$ value in memory.
This leads to a linear memory complexity in terms of time and number of states which would be prohibitive given the full search graph.

Instead, while doing the forward pass we only store a selected few checkpoints to memory and immediately discard the other $\alpha$ values. During the backwards pass these checkpoints are loaded and only the $\alpha$ values for this specific block are recomputed and stored to memory.
If we choose the block size to be $\sqrt{T}$ frames, then the memory consumption goes down to $\mathcal{O}\left(S\cdot 2 \sqrt{T}\right)$ at the expense of having to do the forward pass twice.

This reduction is usually sufficient to handle all training samples of reasonable length. 
If a handling of longer sequences is necessary we propose a more complicated scheme with multiple layers of checkpointing where for each sub-block we again use checkpoints within the block to further reduce memory consumption.
Ideally, if in each iteration the block size is reduced by a factor $\frac{1}{2}$, we arrive at a scheme with logarithmic memory consumption.
The disadvantage is, that in this case the $\alpha$ values have to be reconstructed and discarded more often so that the time requirement is increased by a factor $\log(T)$.

Further savings can be obtained by pruning away states with low occupation probability and only checkpointing a reduced number of states to memory \cite{Zweig2000ExactAC}.
But this approach would then reintroduce restrictions on the sums as in the lattice-based case and is therefore not applied.

An overview of the proposed methods can be seen in Table~\ref{tab:complexity_of_methods}.
For our experiments we used equidistant checkpoints if not noted otherwise.

\begin{table}[th]
  \caption{Memory and runtime complexity of proposed algorithms in terms of total number of frames $T$, number of states $S$ and edges $E$ in the decoding graph }
  \label{tab:complexity_of_methods}
  \centering
  \begin{tabular}{ l | c | c }
    \toprule
    Checkpoints & Runtime  & Memory  \\
    \midrule
    none & $E\cdot T$ & $S\cdot T$ \\
    equidistant & $E\cdot T$ & $S\cdot \sqrt{T}$ \\
    logarithmic & $E\cdot T\log(T)$ & $S\cdot \log(T)$\\
    \bottomrule
  \end{tabular}
  
\end{table}

\section{Experimental setup}
\label{sec:exp_setup}
For our experiments we used the Switchboard training data.
Since our cross-entropy baseline on this task has already strongly been optimized, we also report results on data collected in the Quaero project \cite{quaero}.
The size of the training data is comparable.
We use the 300h portion of the original Switchboard corpus and 250h of broadcast news and pod-casts for Quaero

The acoustic model is a 6-layer bidirectional LSTM with 500 cells per layer and direction.
We use 4500/9000 clustered triphone states as targets (Quaero/Switchboard).
We use L2 normalization, dropout \cite{srivastava2014dropout}, gradient noise \cite{neelakantan2015adding}, and focal loss \cite{lin2017focal} as regularization techniques.
No speaker adaptation or data perturbation techniques were used.
The models are first trained until convergence with the cross-entropy (CE) objective function.
The CE Model is used to generate the lattices for lattice-based sequence discriminative training.
Both lattice-based and lattice-free training is initialized with the CE Model and the learning rate is chosen to be smaller than in the CE training.

For lattice-free training, the search space is constructed as a static composition of a word-level unigram language model, the training lexicon, triphone context with across word coarticulation, and the 3-state HMM topology. 
The final size of the search space automaton is many times larger compared to the previous approach \cite{Povey2016LFMMI}. 
Sizes are given in Table \ref{tab:automata}.

\begin{table}[t]
  \caption{Number of states and edges in the decoding graph for different corpora}
  \label{tab:automata}
  \centering
  \begin{tabular}{ l | c | c }
    \toprule
    Corpus & States  & Edges  \\
    \midrule
    Switchboard \cite{Povey2016LFMMI} &  24k &  220k  \\
    Switchboard                      & 185k &  660k  \\
    Quaero                           & 375k & 1.25M \\
    Librispeech                      & 550k & 2.5M  \\
    \bottomrule
  \end{tabular}
  \vspace{-5mm}
\end{table}

Decoding is done with a count-based fourgram language model.

\section{Results}
\label{sec:results}
\subsection{Denominator Language Model}
An important difference between lattice-free and lattice-based sequence training is the language model used in training. 
We would like to compare the influence of different language model context lengths on lattice-based sequence training performance.
Tables \ref{tab:lattice_lm_swb} and \ref{tab:lattice_lm_qro} show the results for lattice-based MMI and sMBR training for different language models used in comparison with the best lattice-free result.

For the Quaero dataset, we see that for best performance a bigram language model should be used. 
This is in accordance to previous results \cite{Schluter99}.
For Switchboard we only saw improvements if a fourgram language model was used and even degradations if a unigram was used instead.

\begin{table}[bh]
  \caption{Influence of language model context on lattice-based sequence training performance on the Switchboard 300h dataset. Numbers are WER [\%]. }
  \label{tab:lattice_lm_swb}
  \centering
  \begin{tabular}{ c | c | c | c | c }
    \toprule
    Criterion & LM train & LM lattice & Total & SWB   \\
    \midrule
    CE  &  -    & -     & 14.4 & 9.9 \\
    \midrule
    MMI & 1gram & 1gram &  15.2 & 10.3  \\
        & 1gram & 4gram &  15.4 & 10.5  \\
        & 4gram & 4gram &  14.3 & 9.6  \\
    lattice-free & 1gram & - &14.2  & 9.9 \\
    \midrule
    sMBR & 1gram & 1gram & 14.4 & 9.7 \\
         & 1gram & 4gram & 14.2 & 9.4 \\
         & 4gram & 4gram & 13.9 & 9.5 \\
    lattice-free & 1gram & - & 14.3 & 10.0 \\
    \bottomrule
  \end{tabular}
  \vspace{-5mm}
\end{table}

\begin{table}[bh]
  \caption{Influence of language model context on lattice-based sequence training performance on the Quaero dataset. Numbers are WER [\%]. }
  \label{tab:lattice_lm_qro}
  \centering
  \begin{tabular}{ c | c | c | c | c }
    \toprule
    Criterion & LM train & LM lattice & dev'10 & eval'13   \\
    \midrule
    CE  &  -    & -     & 14.2 & 10.7 \\
    \midrule
    MMI & 1gram & 1gram & 13.8  & 10.3  \\
        & 2gram & 2gram & 13.4  & 10.2  \\
        & 4gram & 4gram & 13.9  & 10.5  \\
   lattice-free & 1gram & - &13.8  & 10.3 \\
    \midrule
    sMBR & 1gram & 1gram & 13.5 & 10.2 \\
         & 2gram & 2gram & 13.2 & 10.0 \\
         & 4gram & 4gram & 13.7 & 10.4 \\
    lattice-free & 1gram & - &13.8  & 10.4 \\
    \bottomrule
  \end{tabular}
\vspace{-5mm}
\end{table}

\pagebreak
An advantage of lattice-free sequence discriminative training is the reduced training time and disk space footprint. 
The disk space argument is certainly true, the lattices take up several gigabytes of storage while the search space automaton only needs a few ten megabytes and can be generated on the fly.

\subsection{Runtime Performance}
We now compare the runtime performance of lattice-based and lattice-free sequence training with the baseline cross-entropy training in Table \ref{tab:timing}.
For a fair comparison we have to additionally include the time it takes to create and post-process the lattices in the timing information.
We decided to not do it here as these calculations are trivially parallelizable over many cheap CPU cores, so if it is necessary to quickly train and deploy a model, this task may only take a fraction of overall training time.
Also, in frame-wise cross entropy training we have the possibility to split all utterances into equally sized chunks of small length (e.g. 50 frames) to further speed up the computation.
This is not realistic for sequence criteria without imposing strong approximations on the criterion \cite{Povey2016LFMMI}.
We therefore give two numbers for cross-entropy, one with and one without usage of the chunking technique.

\begin{table}[bh]
  \caption{Training time and real time factor (rtf) per full epoch on the Switchboard data set for different training criteria}
  \label{tab:timing}
  \centering
  \begin{tabular}{ l | c | c  }
    \toprule
    Criterion & Time [h] & RTF   \\
    \midrule
    CE  & 3.5 & 0.012 \\
    -- chunking & 5.0 & 0.017\\
    \midrule
    Lattice-based MMI & 35.0 & 0.12 \\
    Lattice-free MMI & 34.2 &  0.11\\
    \bottomrule
  \end{tabular}
\end{table}

We found a tenfold increase in training times per epoch independent of the method used compared to the cross-entropy baseline. 
Interestingly, there is already an increase by factor $1.5$ when going from a chunked to a non-chunked training procedure.
The time used for the numerator and denominator calculations is about $85\%$ of the total training time.
This is in strong contrast to the results reported in \cite{Povey2016LFMMI}, where these calculations only took up a small fraction of the total training time.
But this discrepancy is to be expected, as the training time is proportional to the number of edges in the search graph and as our implementation does the forward computations twice (cf. Section \ref{sec:criteria}).

This is, however, not a severe limitation of our approach, since we use sequence discriminative training only as a fine-tuning step after successfully training a cross-entropy model.
The total time spent for discriminative training of the neural network is still comparable to the frame-wise training time.

\subsection{Convergence Speed}
\begin{figure}[t]
\begin{tikzpicture}
\begin{axis}[
    width=0.45\textwidth,
    height=0.3\textwidth,
    ylabel={Word Error Rate [\%]},
    xlabel={Number of training epochs},
    xmin=0, xmax=2.0,
    ymin=13.5, ymax=15.0,
    legend pos=north east,
    ymajorgrids=true,
    grid style=dashed,
]
 
\addplot[color=red,mark=None,]
    coordinates {(0,14.216188524590164)(2,14.216188524590164)};

\addplot[color=blue,mark=triangle,]
    coordinates {
(0.0, 14.216188524590164)
(0.25, 13.763661202185792)
(0.5, 13.649817850637522)
(1.0, 13.655510018214935)
(2.0, 13.834813296903461)
    };
    
\addplot[color=green,mark=square,]
    coordinates {
    (0.0, 14.216188524590164)
	(0.16666666666666666, 14.89071038251366)
	(0.5, 14.162112932604735)
	(1.0, 13.840505464480874)
	(1.5, 13.763661202185792)
	(2.0, 13.769353369763206)
    };

\legend{CE Baseline, Lattice-Based MMI, Lattice-Free MMI}
\end{axis}
\end{tikzpicture}
	\caption{Convergence speed of lattice-free and lattice-based MMI training on Quaero dataset, both using unigram word LM. WER on the dev2010 set.}
	\label{fig:convergence}
	\vspace{-5mm}
	\end{figure}
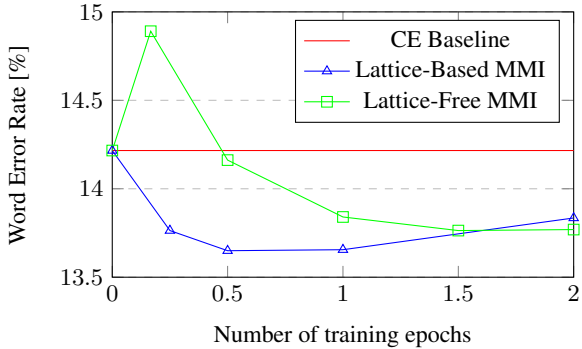

We also want to compare lattice-based and lattice-free methods in terms of convergence speed. 
In Figure \ref{fig:convergence} the progression of word error rate over two epochs of training is shown.

We observe that the lattice-based curve not only converges faster to a minimum but also achieves slightly better error rates given the same conditions.
For the lattice-based case, we also confirmed previous observations \cite{microsoft-2017-conversational-speech-recognition-system,Vesel2013SeqTr} that most of the improvements from sequence discriminative training happen in a fraction of an epoch.
The lattice-free curve shows first an increased error rate which then goes down below the cross-entropy starting point.
This might indicate that the lattice-free model leaves the vicinity of a local optimum around the CE-model. 
It does, however, not lead to a lower error rate.

\subsection{Silence}
\vspace{-1mm}
An important observation of our experiments was that silence requires far more attention in the lattice-free case than if lattices are used.

It is common practice to introduce a scale to each probability model to separately broaden or sharpen the distribution and to tune the relative weight of each model separately.
When we use Viterbi approximation, only the relative scales matter.
For sequence training and when using the sum also the value of the global scale matters.

In lattice-based sequence training it has empirically been found that setting the language model scale to $1$ and reducing the acoustic scale by a factor on theorder of the best LM scale used in recognition gives generally good performance \cite{Vesel2013SeqTr,Schluter99}.
For lattice-free training, this choice leads to a serious degradation in performance as more and more frames are classified as silence frames.
Figure \ref{fig:sil_ratio} shows how the silence ratio changes over the course of the training.

For lattice-based training the ratio stays constant.
In the lattice-free case, the ratio increases depending on the scales.
The best performance for lattice-free training was achieved by not scaling the models, i.e. setting all scales to $1$.

\subsubsection{Silence weight for sMBR}
\vspace{-1mm}
In our experiments with lattice-free sMBR criterion we noticed a severe degradation compared with the lattice-based case. 
In \cite{Kanda2018LatticefreeSM} it was suggested to downweight silence frames in the accuracy computation by a constant factor.
We also tried this (Table \ref{tab:sil_weight}) but found that this weighting did not influence the result.

\begin{table}[bh]
\vspace{-2mm}
  \caption{Influence of weighting silence frame accuracy by a constant factor for lattice-free sMBR training. Numbers represent WER [\%]. }
  \label{tab:sil_weight}
  \centering
  \begin{tabular}{ c | c | c  }
    \toprule
    Silence weight & Total & SWB   \\
    \midrule
	$10^0$    & 14.3 & 10.0 \\
	$10^{-1}$ & 14.3 & 10.0 \\
	$10^{-2}$ & 14.3 & 10.0 \\
	$10^{-3}$ & 14.4 & 10.0 \\
    \bottomrule
  \end{tabular}
  		\vspace{-5mm}
\end{table}

\subsubsection{Acoustic and Language Model scales}
\begin{figure}[t]
\vspace{2mm}
\begin{tikzpicture}
\begin{axis}[
    width=0.45\textwidth,
    height=0.3\textwidth,
    ylabel={Silence Ratio},
    xlabel={Number of training epochs},
    xmin=0, xmax=2.0,
    ymin=0, ymax=0.5,
    legend pos=north west,
    ymajorgrids=true,
    grid style=dashed,
]
 
\addplot[color=red,mark=None,]
    coordinates {(0,0.048924964245)(2,0.048924964245)};

\addplot[color=blue,mark=triangle,]
    coordinates {
    (0,0.048924964245)
(0.25, 0.050316159049404681)
(0.5, 0.048590487387564954)
(0.75, 0.049166120442310848)
(1.0, 0.047182982106264072)
(1.25, 0.047577006080225075)
(1.5, 0.047696402176581693)
(1.75, 0.047830575455276129)
(2.0, 0.0472748048)
    };
    
\addplot[color=green,mark=square,]
    coordinates {
	(0,0.048924964245)
(0.16666666666666666, 0.088063364918803927)
(0.33333333333333331, 0.089878775435120864)
(0.5, 0.093086874416436399)
(0.66666666666666663, 0.099887212052876456)
(0.83333333333333337, 0.11852734217124501)
(1.0, 0.34005431606457487)
(1.1666666666666667, 0.38696220356818484)
(1.3333333333333333, 0.38888794556002715)
(1.5, 0.39040775379273734)
(1.6666666666666667, 0.38328750816767759)
(1.8333333333333333, 0.39238167523532103)
(2.0, 0.39227426482535643)
    };

\addplot[color=green,mark=*]
    coordinates {
    (0,0.048924964245)
(0.16666666666666666, 0.099029688199477994)
(0.33333333333333331, 0.088083127218565677)
(0.5, 0.084498416875136539)
(0.66666666666666663, 0.082869227168026346)
(0.83333333333333337, 0.080396592623501073)
(1.0, 0.073532631742708437)
(1.1666666666666667, 0.079163799941219568)
(1.3333333333333333, 0.079240700421128199)
(1.5, 0.079715142301191685)
(1.6666666666666667, 0.078164452217913932)
(1.8333333333333333, 0.077836905106884333)
(2.0, 0.069235804922166411)
    
    };

\legend{CE,LB, LF AM:0.1, LF AM:1.0}
\end{axis}
\end{tikzpicture}
	\caption{Fraction of the training frames classified as silence over time for the cross-entropy (CE) baseline and lattice-based (LB) and lattice-free (LF) MMI trained models using different acoustic model scales (AM). }
	\label{fig:sil_ratio}
		\vspace{-5mm}
	\end{figure}
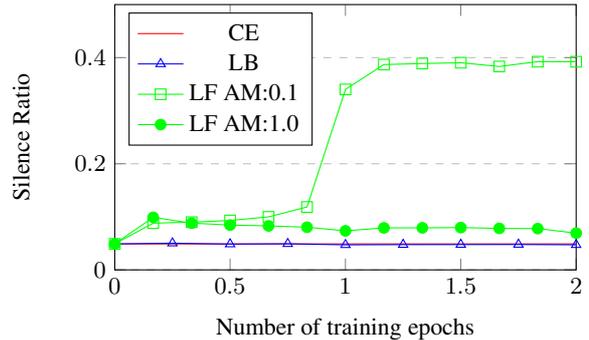

\vspace{-1mm}
As already shown in Figure \ref{fig:sil_ratio} the scales of our models used during training can have a strong influence on the convergence of lattice-free training.
In Table \ref{tab:scales} we present our results.

While a low acoustic scale can seriously degrade the performance of the trained model, if it falls under a certain threshold, the language model scale seems to be of minor influence.

\begin{table}[bh]
  \caption{Influence of acoustic (AM) and language model (LM) scale on training performance for lattice-free MMI training. The other scale has been kept constant at $1.0$. Numbers represent WER [\%] on the total Hub5'00 set. }
  \label{tab:scales}
  \centering
  \begin{tabular}{ c | c  }
    \toprule
    AM scale & Total    \\
    \midrule
    0.1    & 26.0  \\
	0.2    & 14.7  \\
	1.0    & 14.2  \\
    \bottomrule
  \end{tabular}
  \qquad
  \begin{tabular}{ c | c  }
    \toprule
    LM scale & Total    \\
    \midrule
    0.2    & 14.2  \\
	1.0    & 14.2  \\
	10.0 & 14.3  \\
    \bottomrule
  \end{tabular}
		\vspace{-5mm}
\end{table}

\section{Conclusions}
\label{sec:conclusions}
In this work we presented a memory-efficient approach for calculating lattice-free sequence training criteria on GPU despite using a full word language model and full length utterances.

Comparing lattice-based and lattice-free training criteria we found that with careful tuning the lattice-free MMI trained model approaches the performance of a comparable lattice-based MMI model.
The Lattice-based sMBR criterion still outperforms all lattice-free training criteria while the lattice-free version consistently performed worst.

Special attention is required with respect to silence frames in the lattice-free case.
Without restrictions on set word-boundaries silence tends to dominate the model.
Similar problems have been encountered in the case of frequent realignments using LSTM acoustic models in the context of full-sum training or CTC \cite{zeyer2017:ctc}.

\vspace{-1mm}
\begin{center}
{\bf \large Acknowledgements\vspace{-.5em}\vspace{2pt}}
\end{center}\begingroup
\begin{small}
\setlength{\columnsep}{2pt}%
\setlength{\intextsep}{0pt}%
This work has received funding from the European Research Council (ERC) under the European Union's Horizon 2020 research and innovation programme (grant agreement No 694537, project "SEQCLAS" and Marie Skłodowska-Curie grant agreement No 644283, project "LISTEN") and from a Google Focused Award. 
The work reflects only the authors' views and none of the funding parties is responsible for any use that may be made of the information it contains. 
The GPU cluster used for the experiments was partially funded by Deutsche Forschungsgemeinschaft (DFG) Grant INST 222/1168-1. 

The authors would like to thank Albert Zeyer, Eugen Beck, and Wei Zhou for many helpful discussions.
\end{small}
\endgroup

\bibliographystyle{IEEEtran}

\bibliography{refs}

\end{document}